\documentclass[11pt,a4papers,floats]{article}
\usepackage{amsfonts,color,epsfig}
\usepackage{multirow}
\usepackage[multiple]{footmisc}
\usepackage{booktabs}
\newcommand{\nc}{\newcommand}

\renewcommand{\thefootnote}{\fnsymbol{footnote}}
\nc{\fig}[5] {
\begin{figure}[!htbp]
    \begin{center}
    \leavevmode
    \centerline{
        \includegraphics[width=#1, height=#2]{#3}
        }
    \caption[]{#4}
    \label{#5}
    \end{center}
\end{figure}}
\nc{\figs}[8]{
\begin{figure}[!htbp]
    \begin{center}
    \leavevmode
    \centerline{
        \includegraphics[width=#1, height=#2]{#3}
        \includegraphics[width=#4, height=#5]{#6}
        }
    \caption[]{#7}
    \label{#8}
    \end{center}
\end{figure}}

\begin{document}
\begin{flushright}
{\small \tt arXiv:0904.0085\\
\tt [hep-th]}
\end{flushright}
\vspace{2mm}
\begin{center}
{\Large {\bf Absorption cross section in the
    topologically massive gravity at the critical point}}\\[5mm]
{John J. Oh$^{a}$\footnote{Email: johnoh@nims.re.kr} and Wontae Kim$^{b,c}$\footnote{Email: wtkim@mail.sogang.ac.kr}}\\[5mm]
{\small {${}^{a}$\it Division of Interdisciplinary Mathematics, 
National Institute for Mathematical Sciences,\\ Daejeon 305-340, South Korea\\[0pt]
${}^{b}$ \it {Department of Physics, Sogang University, C.P.O. Box 1142, Seoul 100-611, South Korea} \\[0pt]
${}^{c}$ \it Center for Quantum Spacetime, Sogang University, Seoul 121-742, South Korea      
}}
\end{center}
\vspace{1mm}
\begin{abstract}
The absorption cross section for 
the the warped AdS$_3$ black hole background shows that it
is larger than the area even 
if the s-wave limit is considered.
It raises some question whether the deviation from the areal cross section is due to 
the warped configuration of the geometry or the rotating coordinate
system, where these two effects are mixed up in the warped AdS$_3$ black hole. 
So, we study the low-frequency scattering dynamics of propagating
scalar fields 
under the warped AdS$_3$ background at the critical point 
which reduces to the BTZ black hole in the
rotating frame without the warped factor, which 
shows that the deformation effect at the critical point 
does not appear. 
\end{abstract}
\vspace{2mm}


\vspace{3mm}

\hspace{10.5cm}{Typeset Using \LaTeX}
\newpage
\renewcommand{\thefootnote}{\arabic{footnote}}
\setcounter{footnote}{0}

There has been much attention to three-dimensional gravities 
since it might be a potential test bed to quantum gravity.
In the Einstein-Hilbert action with 
a cosmological constant, there are no propagating degrees of freedom
in the bulk 
even though there is a asymptotically AdS$_3$ black hole solution
called the Bandos-Teitelboim-Zanelli (BTZ) black 
hole \cite{Banados:1992wn}. 
Moreover, if we consider
the gravitational Chern-Simons (GCS) term \cite{Deser:1981wh} additionally, 
there has a single massive graviton mode propagating in the bulk,
where the BTZ black hole as a trivial class of solutions \cite{Banados:1992gq}
leads to unstable vacua of AdS$_3$ black holes.
Recently, there have been many intensive studies on the issue 
of ``chiral gravity'' and ``missing gravitons'' 
\cite{Li:2008dq,Grumiller:2008qz,Giribet:2008bw,Grumiller:2008es,Carlip:2008jk,Li:2008yz,Carlip:2008eq,Sachs:2008gt,Grumiller:2008pr,Strominger:2008dp,Carlip:2008qh,Park:2008yy}.

The three-dimensional Einstein gravity 
with a negative cosmological constant $\Lambda=-2/\ell^2$ and the GCS term with the coupling
constant of $1/\mu$ \cite{Li:2008dq} is defined by
\begin{equation}
\label{eq:tmg}
S=\frac{1}{2\kappa^2} \int d^3 x \sqrt{-g}
\left(R+\frac{2}{\ell^2}\right)+ \frac{1}{4\kappa^2\mu} \int d^3 x
\sqrt{-g} \tilde{\epsilon}^{\lambda\mu\nu} \Gamma_{\lambda\sigma}^{\rho}\left[\partial_{\mu} \Gamma_{\nu\rho}^{\sigma} + \frac{2}{3}\Gamma_{\mu\tau}^{\sigma}\Gamma_{\nu\rho}^{\tau}\right],
\end{equation}
where $\kappa^2=8\pi G_{N}$ associated with the three-dimensional
Newton's constant $G_{N}$, $\tilde{\epsilon}^{\lambda\mu\nu}$ is a
tensor defined 
by ${\epsilon}^{\lambda\mu\nu}/\sqrt{-g}$ with ${\epsilon}^{012}=1$, 
and $\mu\equiv 3\nu/\ell$ is the dimensionless coupling constant. If
we choose a positive sign for the Einstein-Hilbert term as above, 
the black hole has a positive energy for $\mu\ell >1$ while massive
gravitons have the negative energy. 
Apart from the unstable solution, the ``warped AdS$_3$ vacua'' has
been proposed as a fibration of the real line with a constant warp
factor over AdS$_2$ geometry, which reduces the $SL(2,R)_{L}\times
SL(2,L)_{R}$ isometry group to $SL(2,R)\times U(1)$
\cite{Anninos:2008fx}. 
Among the solutions of the cosmological topologically massive gravity
(CTMG), the only closed time-like curve(CTC)-free one 
is known as the spacelike stretched black hole (simply called ``warped
AdS$_3$ black hole''), 
which has been also studied in Refs. \cite{Bouchareb:2007yx,Moussa:2008sj}.

One of the most intriguing properties of the
warped black hole is that the whole spacetime is rotating
in the sense that 
the angular velocity of the geometry cannot vanish even in spite of
the vanishing angular momentum of the warped black hole, 
which is drastically different from the standard BTZ black
hole. It means the whole spacetime is nothing but the ergoregion so
that the geometry of the warped AdS$_3$ black hole looks like inside 
the ergoregion of a rotating black hole. In other words, there is no
static observer in our spacetime except the infinity \cite{Kim:2008bf}.
This property of the geometry is valid for $\nu \ge 1 $ in this black hole.

Recently, there has been the study of the behavior of a scalar field
on the spacelike stretched black
hole background and its absorption cross section 
using the low-frequency scattering procedure \cite{Oh:2008tc}, 
in which the absorption cross section in the s-wave sector 
of propagating scalar fields under the warped AdS$_3$ black hole 
for $\nu\ne 1$ is given by
\begin{equation}
\label{eq:absT}
\sigma_{abs}^{\mu=0} = {\mathcal A}_{H} + d(\nu,T_R, T_L),
\end{equation}
where ${\mathcal A}_{H}$ is the area at the horizon and  $d(\nu,T_R,T_L)$ is the positive deformation
factor depending on the warped factor with the left-handed and right-handed temperatures. It
shows that the absorption cross section is larger than the expected
value of area. 
However, it is still unclear whether the origin of this large
cross section is due to the warped effect of the geometry or the
rotating effect from the entire ergoregion mentioned earlier.

So, we would like to study what the relevant origin to this deformation of the absorption
cross section is for this warped AdS$_3$ black hole. 
For this purpose, we will exclude the warped effect from the warped
AdS$_3$ black hole, which will be realized simply by setting $\nu =1$
condition to the warped metric so that only the rotating effect remains.   
Then, the resulting metric is just the BTZ solution in the rotating
frame; however, the ergoregion is maintained in that the whole spacetime
is still ergoregion even in this limit.
It is noteworthy that,
in the BTZ limit in the warped black hole and the standard BTZ
solution in the CTMG, 
the former case still does not have a vanishing angular velocity at all, 
while the latter case can have a vanishing angular velocity
limit. They are related by the coordinate transformation
which makes the latter spacetime outside the horizon into the totally
ergoregion. 

In this paper, 
we shall investigate the absorption cross section 
for a propagating scalar field in the background of the entirely
rotating geometry which corresponds to the critical case of $\nu=1$ for the warped
AdS$_3$ black hole. Actually, this critical case is ill-defined in the
above formula (\ref{eq:absT}) so that we have to calculate relevant equations 
by setting $\nu=1$ from the beginning. As a consequence, 
we will find the areal scattering cross section, which implies
that the warped effect than the rotating effect 
is responsible for the deformation of the absorption cross section. 

Now, varying the action (\ref{eq:tmg}) with respect to the metric 
leads to the bulk equation of motion,
\begin{equation}
\label{eq:eqnmot}
G_{\mu\nu} - \frac{1}{\ell^2}g_{\mu\nu} + \frac{\ell}{3\nu} C_{\mu\nu}=0,
\end{equation}
where the Cotton tensor is
\begin{equation}
C_{\mu\nu} = \epsilon_{\mu}^{~\lambda\sigma}\nabla_{\lambda}\left(R_{\sigma\nu} - \frac{1}{4} g_{\sigma\nu} R\right).
\end{equation}
The warped AdS$_3$ black hole solution is given by 
 \cite{Anninos:2008fx}, 
\begin{equation}
\label{eq:met}
(ds)^2 = -N^2(r) dt^2 + \ell^2 R^2(r) (d\theta + N^{\theta}(r)dt)^2 + \frac{\ell^4 dr^2}{4R^2(r)N^2(r)},
\end{equation}
for $\nu^2>1$
where the metric functions are
\begin{eqnarray}
&& R^2(r) = \frac{r}{4}\left(3(\nu^2-1)r+(\nu^2+3)(r_{+}+r_{-})-4\nu\sqrt{r_{+}r_{-}(\nu^2+3)}\right),\\
&& N^2(r) = \frac{\ell^2(\nu^2+3)(r-r_{+})(r-r_{-})}{4R^2(r)},\\
&& N^{\theta}(r) = \frac{2\nu r-\sqrt{r_{+}r_{-}(\nu^2+3)}}{2R^2(r)}.
\end{eqnarray}
They were already discovered in Refs. \cite{Israel:2004vv,Detournay:2005fz}
as a marginal deformation of the $SL(2,R)$ Wess-Zumino-Witten model in the context
of string theory and discussed the connection with
the warped solution \cite{Compere:2008cw} and the dual CFT description
of TMG with the warped boundary conditions \cite{Compere:2008cv}  in
Ref.  \cite{Anninos:2008fx}. 
It is closely related to the solution
{originally} discovered in Refs. \cite{Bouchareb:2007yx, Moussa:2008sj}, which is connected with
the coordinate transformation that breaks down at the critical point of $\nu = 1$ and the negative
  $\nu$ yields some unphysical results. 

Now, we are going to pay attention to the critical point of
$\nu=1$ so that the metric functions should be reduced to
\begin{equation}
R^2(r) = r\left(\sqrt{r_{+}}-\sqrt{r_{-}}\right)^2,~~ N^2(r) = \frac{\ell^2(r-r_{+})(r-r_{-})}{r(\sqrt{r_{+}}-\sqrt{r_{-}})^2},~~
N^{\theta}(r) = \frac{r-\sqrt{r_{+}r_{-}}}{r(\sqrt{r_{+}}-\sqrt{r_{-}})^2}.
\end{equation}
They can be expressed in terms of the standard form of the BTZ black hole by the coordinate transformation \cite{Kim:2008bf},
\begin{equation}
\label{eq:cotrans}
t=\frac{\rho_{+}-\rho_{-}}{\ell}\tau,~~\theta = \varphi - \frac{1}{\ell}\tau, ~~r=\frac{\rho^2}{\rho_{+}-\rho_{-}}.
\end{equation}
Using this coordinate transformation, we can easily show 
that the thermodynamic internal energy of scalar fields on this
metric (\ref{eq:met}) is not the same with that of the BTZ black hole
background while the entropy and the angular momentum are identical to that of the BTZ
black hole, respectively.
The Killing vectors are defined as $\chi^{a} \equiv \xi^{a} + \Omega_{H}\phi^{a}$ where the angular velocity is 
\begin{equation}
\Omega_{H} \equiv - \left.\frac{g_{t\theta}}{g_{\theta\theta}}\right|_{H} = -\frac{1}{\sqrt{r_{+}}(\sqrt{r_{+}}-\sqrt{r_{-}})},
\end{equation}
where the subscript $H$ denotes the value at the horizon of $r=r_{+}$.
Then, the Hawking temperature can be found from the surface gravity defined as $\kappa_{H}^2 = - \frac{1}{2}(\nabla_{a}\chi_{b})(\nabla^{a}\chi^{b})|_{H}$, which is
\begin{equation}
T_{H} \equiv \frac{\kappa_{H}}{2\pi} = \frac{1}{2\pi}\left(1+\sqrt{\frac{r_{-}}{r_{+}}}\right),
\end{equation}
which is related to those of the BTZ black hole, $\Omega_{H}=[\ell/(\rho_{+}-\rho_{-})]\left(\Omega_{H}^{BTZ}-1/\ell\right)$ and $T_{H}=[\ell/(\rho_{+}-\rho_{-})]T_{H}^{BTZ}$, respectively.

On the other hand, as 
shown in Ref. \cite{Oh:2008tc}, the deformation factor in
Eq. (\ref{eq:absT}) can be computed as
\begin{equation}
d(\nu,T_R, T_L)=\frac{64\pi^2\nu^3(\nu^2+2)(\nu^2+4)}{3(\nu^2+3)(\nu^2-1)}\left[1 + \frac{1}{4\nu}\sqrt{(\nu^2+3)\left(1-\frac{3(\nu^2+3)(\nu^2-1)T_{R}}{8\pi\nu^2(T_{L}+T_{R})^2}\right)}\right].
\end{equation}
Note that the factor diverges when $\nu\rightarrow 1$, and the analysis in
Ref. \cite{Oh:2008tc} is no more valid for this case. The
reason for this is that the functional
transformation of the hypergeometric function is singular at $\nu=1$,
which implies that the matching procedure and the corresponding result
can be drastically 
different from the case of $\nu\ne 1$. 
So, for $\nu=1$ which is free from the warped effect, 
we have to study whether the deformation can survive
or not. As was mentioned, the BTZ black hole in the rotating frame
is different from the original one in that the ergoregion is the whole
spacetime so that the static observer does not exist.   


Now, let us start with the Klein-Gordon equation of the massless scalar field
on the spacelike stretched warped AdS$_3$ black hole background,
\begin{equation}
\frac{1}{\sqrt{-g}}\partial_{\mu}\left(\sqrt{-g}g^{\mu\nu}\partial_{\nu}\Phi\right) = 0,
\end{equation}
where $\Phi=\Phi(t,r,\theta)$. 
Using the separation of variables of $\Phi(t,r,\theta)\equiv
\Phi(r)e^{-i\omega t+i\mu\theta}$, the radial equation of motion can be written as
\begin{eqnarray}
\label{eq:radeqn}
 \Phi''(r) + \frac{(2r-r_{+}-r_{-})}{(r-r_{+})(r-r_{-})}\Phi'(r) -\frac{\beta r + \gamma}{(r-r_{+})^2(r-r_{-})^2}\Phi(r)=0,
\end{eqnarray}
where
\begin{eqnarray}
&& \beta \equiv -\frac{\omega^2(\sqrt{r_{+}}-\sqrt{r_{-}})^2+2\mu\omega}{4},\label{eq:para2}\\
&& \gamma \equiv -\frac{\mu(\mu-2\omega\sqrt{r_{+}r_{-}})}{4}\label{eq:para3}
\end{eqnarray}
and the prime denotes a derivative with respect to $r$.
The general solution is found in terms of the second kind hypergeometric functions,
\begin{eqnarray}
\Phi(r) &=& C_{1} \frac{\left(r-r_{+}\right)^{\alpha_{+}}}{\left(r-r_{-}\right)^{-\alpha_{-}}} F\left(\alpha_{+}-\alpha_{-}, 1+\alpha_{+}-\alpha_{-}, 1+2\alpha_{+}; \frac{r_{+}-r}{r_{+}-r_{-}}\right)\nonumber\\
&+& C_{2} \frac{\left(r-r_{+}\right)^{-\alpha_{+}}}{\left(r-r_{-}\right)^{\alpha_{-}}} F\left(-\alpha_{+}-\alpha_{-}, 1-\alpha_{+}-\alpha_{-}, 1-2\alpha_{+}; \frac{r_{+}-r}{r_{+}-r_{-}}\right)\label{eq:hgfsol}
\end{eqnarray}
where $\alpha_{\pm} \equiv \frac{\sqrt{\beta r_{\pm}+\gamma}}{r_{+} - r_{-}}$.

If we consider the near-horizon limit of $r\simeq r_{+}$, the general solution becomes
\begin{eqnarray}
\label{eq:nearsol}
\Phi_{near}(r) &\simeq& \hat{C}_{1} \left(r-r_{+}\right)^{\alpha_{+}} + \hat{C}_{2} \left(r-r_{+}\right)^{-\alpha_{-}}\nonumber\\
&=& \hat{C}_{1} {\rm exp}\left[\alpha_{+} \ln (r-r_{+})\right]+\hat{C}_{2} {\rm exp}\left[-\alpha_{-}\ln (r-r_{+})\right],
\end{eqnarray}
where $\hat{C}_{1,2} \equiv
C_{1,2}\left(r_{+}-r_{-}\right)^{-\alpha_{-}}$. Note that in the low-frequency
limit of $\omega << 1$, we have the purely imaginary in the exponent since
\begin{equation}
\sqrt{\beta r_{+} +\gamma} = i\frac{1}{2}\left[\mu+(r_{+}-\sqrt{r_{+}r_{-}})\omega\right]+ {\mathcal O}(\omega^2),
\end{equation}
which gives the ``in-going'' and ``out-going'' coefficients, $C_{in} \equiv {C}_{2}$ and $C_{out} \equiv {C}_{1}$, respectively. Now, we need to impose a boundary condition from the corresponding physical situation. In general, two-independent boundary conditions can be imposed from the two equivalent pictures of probing scalar fields under the black hole background. The first one is for the classical description of black holes; a black hole can absorb the probing fields but nothing can escape from the event horizon, implying that the out-going coefficient near the horizon should vanish. The second one is for the quantum-mechanical description of the black hole; the asymptotic observer can see the quantum radiation coming from the black hole horizon, implying that the in-going coefficient at asymptotic region should vanish. Both descriptions are equivalent and independent each other, so we shall use the first one, i.e., $C_{out} = 0$ for convenience.

Here, if we use the transformation rule for the hypergeometric function
in Eq. (\ref{eq:hgfsol}) \cite{as},
\begin{eqnarray}
F(a,a+1;c;z) &=& \frac{\Gamma(c)(-z)^{-a-1}}{\Gamma(a+1)\Gamma(c-a)} \sum_{n=0}^{\infty} \frac{(a)_{n+1} (1-c+a)_{n+1}}{n!(n+1)!} z^{-n} \nonumber\\&\times& \left(\ln(-z)+ \psi(2+n)+\psi(1+n)-\psi(a+1+n)-\psi(c-a-1-n)\right)\nonumber\\
&+&(-z)^{-a} \frac{\Gamma(c)}{\Gamma(a+1)\Gamma(c-a)}
\end{eqnarray}
where $\psi$ is a digamma function and $(a)_n \equiv
a(a+1)(a+2)\cdots(a+n-1)$ with $(a)_0=1$. 
Then, the transformed general solution in the leading order is
\begin{eqnarray}
\Phi_{trans}(r) &=& C_{in} \frac{ \left(r-r_{+}\right)^{\alpha_{-}}}{\left(r-r_{-}\right)^{\alpha_{-}} }\frac{\Gamma(1-2\alpha_{+})\left({r_{+}-r_{-}}\right)^{-\alpha_{+}-\alpha_{-}}}{\Gamma(1-\alpha_{+}-\alpha_{-})\Gamma(1-\alpha_{+}+\alpha_{-})} \left[ 1 - \left(\frac{\beta}{r-r_{+}}\right)\right. \nonumber\\ &\times&  \left.\left\{\ln\left(\frac{r-r_{+}}{r_{+}-r_{-}}\right)+\psi(2)+\psi(1)-\psi(1-\alpha_{+}-\alpha_{-}) - \psi(\alpha_{-}-\alpha_{+})\right\} \right],
\end{eqnarray}
which becomes in the limit of $r\rightarrow \infty$,
\begin{eqnarray}
\label{eq:trans}
\Phi_{trans}^{r\rightarrow\infty}(r) = C_{in} \frac{\Gamma(1-2\alpha_{+})\left({r_{+}-r_{-}}\right)^{-\alpha_{+}-\alpha_{-}}}{\Gamma(1-\alpha_{+}-\alpha_{-})\Gamma(1-\alpha_{+}+\alpha_{-})}\left[ 1 - \frac{\beta}{r}\ln r \right].
\end{eqnarray}

On the other hand, in the asymptotic region of $r\rightarrow \infty$, the wave equation can be written in the form of
\begin{equation}
\Phi''(r) + \frac{2}{r} \Phi'(r) +\frac{\hat{\beta}}{r^3} \Phi(r) = 0,
\end{equation}
whose solution is given by the linear combination of the Bessel functions, $J_{\nu}(x)$ and $Y_{\nu}(x)$,
\begin{eqnarray}
\Phi_{asym}(r) = \frac{A_{1}}{\sqrt{r}}  {J}_{1}\left(2\sqrt{\hat{\beta}r^{-1}}\right)+ \frac{A_{2}}{\sqrt{r}} {Y}_{1}\left(2\sqrt{\hat{\beta}r^{-1}}\right)
\end{eqnarray}
where we define $\hat{\beta}\equiv -\beta$. 
It is found that the Bessel functions can be expressed 
in the form of the following ascending series,
\begin{eqnarray}
J_{\nu} (z) &=& \left(\frac{z}{2}\right)^{\nu} \sum_{k=0}^{\infty} \frac{1}{k! \Gamma(\nu+k+1)}\left(-\frac{z^2}{4}\right)^{k},\nonumber\\
Y_{\nu} (z) &=& - \frac{1}{\pi} \left(\frac{z}{2}\right)^{-\nu} \sum_{k=0}^{\nu-1} \frac{(\nu-k-1)!}{k!} \left(\frac{z^2}{4}\right)^{k} + \frac{2}{\pi}\ln\left(\frac{z}{2}\right) J_{\nu}(z) \nonumber\\
&-& \frac{1}{\pi}\left(\frac{z}{2}\right)^{\nu} \sum_{k=0}^{\infty} \left(\psi(k+1)+\psi(\nu+k+1)\right) \frac{1}{k!(\nu+k)!} \left(-\frac{z^2}{4}\right)^{k}.
\end{eqnarray}
Thus, the asymptotic solution can be written in the polynomial form of
\begin{equation}
\label{eq:asymsol}
\Phi_{asym}(r) \simeq \hat{A}_{1} + \frac{\hat{A}_{2}}{r} \ln r
\end{equation}
where the coefficients are 
\begin{equation}
\hat{A}_{1} \equiv -\frac{A_{2}}{\pi\hat{\beta}^{1/2}},~~\hat{A}_{2}\equiv -\frac{A_{2}\hat{\beta}^{1/2}}{\pi}.
\end{equation}
Comparing Eq. (\ref{eq:trans}) to Eq. (\ref{eq:asymsol}) leads to
\begin{equation}
\hat{A}_{1}=C_{in} \frac{\Gamma(1-2\alpha_{+})\left({r_{+}-r_{-}}\right)^{-\alpha_{+}-\alpha_{-}}}{\Gamma(1-\alpha_{+}-\alpha_{-})\Gamma(1-\alpha_{+}+\alpha_{-})},~~
\hat{A}_{2}=-\hat{A}_{1}\hat{\beta}.
\end{equation}

Here, we decompose the ''in-going'' and ''out-going'' modes in the
 asymptotic solution by defining $\hat{A}_{1} \equiv A_{in} + A_{out}$
 and $\hat{A}_{2} \equiv ih(A_{out}-A_{in})$. Then,
 we can rewrite the asymptotic solution as
\begin{equation}
\Phi_{asymp}(r) \simeq A_{in} \left(1-i\frac{h}{r}\ln r \right) + A_{out} \left(1+i\frac{h}{r} \ln r\right),
\end{equation}
where $h$ is a positive dimensionless  {numerical} constant which will be taken to be independent of the energy $\omega$ \cite{bss,ko}. Note 
that this constant can be chosen so that the absorption cross section is described by the area of the black hole in the low-frequency regime \cite{Oh:2008tc, bss, dgm}. On the other hand, it can be chosen so as to have the usual value of the Hawking temperature \cite{ko} or to make the sum of absorption and reflection coefficients be unity \cite{kko}.
This ambiguity comes from the fact that there exists an arbitrary freedom when we decompose the amplitude of the wave function into in-going and out-going modes. However, this freedom can be chosen as a numerical factor that is independent of the energy of probing fields by appropriate physical situations.
 
 The absorption (${\mathfrak A}$) and the reflection (${\mathfrak R}$) coefficients are defined by the ratio of ``in-going'' and ``out-going'' fluxes as
\begin{equation}
\label{eq:fluxes}
{\mathfrak A} \equiv \left|\frac{{\mathcal F}_{near}^{in}}{{\mathcal F}_{asym}^{in}}\right|,~~{\mathfrak R} \equiv \left|\frac{{\mathcal F}_{asym}^{out}}{{\mathcal F}_{asym}^{in}}\right|,
\end{equation}
respectively. Note that the definition of the flux is given by
\begin{equation}
{\mathcal F} = \frac{2\pi}{i} (r-r_{+})(r-r_{-})\left[\Phi^{*}(r)\frac{\partial}{\partial r} \Phi(r) - \Phi(r)\frac{\partial}{\partial r} \Phi^{*}(r)\right],
\end{equation}
which yields the ''in-going'' flux at asymptotic region and the ''in-going'' flux near the horizon,
\begin{equation}
{\mathcal F}_{asymp}^{in} = - 4\pi h |A_{out}|^2 \ln r_{b},~~{\mathcal F}_{near}^{in} = - 2\pi |C_{in}|^2 (\mu+\omega\sqrt{r_{+}}(\sqrt{r_{+}}-\sqrt{r_{-}})),
\end{equation}
where $r_{b}$ denotes the boundary of AdS space. The reason why we set
the spatial boundary of $r_b$ is due to the fact that the real AdS
boundary is not the same with that in the rotating frame. Indeed as
seen in Eq. (\ref{eq:cotrans}), the radial coordinate $r$ in the
rotating frame linearly increases while the 
radial coordinate of the standard BTZ black hole increases
quadratically. 
So, the boundary $r_b$ in the $r$-coordinate system at the real AdS boundary looks finite, which can be regarded as a constant boundary surface we denoted as $r_b$.

From Eq. (\ref{eq:fluxes}), one finds the absorption coefficient in the s-wave sector,
\begin{equation}
{\mathfrak A}^{\mu=0}  = \frac{\omega{\mathcal A}_{H}}{4\pi h\ln r_{b}}\frac{ |C_{in}|^2}{|A_{out}|^2}.
\end{equation}
More specifically, the amplitudes can be computed in the low-frequency limit as
\begin{equation}
|A_{out}|^2 = - \frac{|A_2|^2 (h^2 + {\beta}^2)}{4\pi^2 h^2{\beta}},
\end{equation}
and
\begin{equation}
|C_{in}|^2 = \left.-\frac{|A_2|^2 {\rm sinh}\pi \hat{\alpha}_{+}}{2\pi \hat{\alpha}_{+} {\rm sinh}\pi (\hat{\alpha}_{+}+\hat{\alpha}_{-}) {\rm sinh}\pi (\hat{\alpha}_{+}-\hat{\alpha}_{-})}\right|_{\omega\rightarrow 0}^{\mu=0} \simeq \frac{2|A_{2}|^2}{\pi^3\omega^2(r_{+}-r_{-})(\sqrt{r_{+}}-\sqrt{r_{-}})^2},
\end{equation}
where $\hat{\alpha}_{\pm} \equiv - i \alpha_{\pm} \in {\mathbf R}$. 
Therefore, the ratio between ''in-going'' and ''out-going'' coefficients becomes
\begin{equation}
\left.\frac{|C_{in}|^2}{|A_{out}|^2}\right|_{\omega\rightarrow 0} \simeq \frac{2}{\pi(r_{+}-r_{-})}
\end{equation}
and the absorption cross section for s-wave modes in the low-frequency limit is found to be
\begin{equation}
\sigma_{abs}^{\mu=0} \equiv \left.\frac{{\mathfrak A}^{\mu=0}}{\omega}\right|_{\omega\rightarrow 0} \simeq {\mathcal A}_{H}
\end{equation}
where the parameter $h$ is chosen to have an area as $h =
(2h(r_{+}-r_{-})\ln r_{b})^{-1}$. Note that the result shows that the
scattering cross section under the warped AdS$_3$ black hole of 
the case of $\nu=1$ is proportional to the area of the event horizon, which is coincident with the usual case of the BTZ black hole \cite{bss}.

We have investigated the absorption cross section of propagating
scalar fields in the limiting case of the warped AdS$_3$ black hole
background called the critical limit of $\nu=1$. 
In the previous study of the warped geometry ($\nu\ne 1$), 
there is no limit of $\nu=1$ since the functional transformation 
of the hypergeometric function is ill-defined, 
which implies that we need to use a different transformation rule in this case.
It also implies that the asymptotic solution should be altered 
and the corresponding ratio between in-going and out-going fluxes
should be 
changed unlike the $\nu\ne 1$ case. 
In conclusion, the deformation of the absorption cross section
for the warped $AdS_3$ black hole is related to the warped geometry than
the rotating coordinate effect, in other words, it is
irrelevant to the ergoregion.

\vspace{10mm}
{\bf Acknowledgment}
\\
{We would like to thank E. J. Son for exciting discussions.
W. Kim was supported by the Korea Science and Engineering Foundation 
(KOSEF) grant funded by the Korea government(MOST) 
(R01-2007-000-20062-0) and the Korea Science and Engineering
Foundation (KOSEF) grant funded by the Korea government(MEST) through the Center for Quantum Spacetime(CQUeST) of Sogang University with grant number R11 - 2005 - 021.
J. J. Oh was supported by the Korea Research Council
of Fundamental Science \& Technology (KRCF).
}


\end{document}